\documentclass[sigconf, anonymous=false]{acmart}

\usepackage{acronym}
\usepackage{xcolor}

\usepackage{caption}
\usepackage{subfig}
\usepackage{numdef}

\newcommand{\specialcell}[2][c]{%
  \begin{tabular}[#1]{@{}c@{}}#2\end{tabular}}

\newcommand{\etc}{etc.}
\newcommand{\ie}{i.e.}
\newcommand{\eg}{e.g.}

\newcommand{\sTime}{7}

\newcommand{\allAnswers}{350}
\newcommand{\percPrereq}{50}
\newcommand{\percPractice}{50}
\newcommand{\notInTeam}{12}
\newcommand{\finalFiltered}{313}
\newcommand{\automltime}{1 hour}

\usepackage[export]{adjustbox}

\newcommand{\hide}[1]{}

\usepackage{xcolor}

\definecolor{alexcolor}{rgb}{0.4,0.6,0.2}
\definecolor{hhcolor}{rgb}{0.2,0.6,0.6}
\definecolor{joostcolor}{rgb}{0.8,0.5,0.3}
\definecolor{koencolor}{rgb}{0.0,0.0,1}
\definecolor{todocolor}{rgb}{0.9,0.1,0.1}
\definecolor{changedcolor}{rgb}{0.42,0.27,0.57}

\AtBeginDocument{%
  \providecommand\BibTeX{{%
    \normalfont B\kern-0.5em{\scshape i\kern-0.25em b}\kern-0.8em\TeX}}}

\setcopyright{acmcopyright}
\copyrightyear{2020}
\acmYear{2020}

\copyrightyear{2020}
\acmYear{2020}
\setcopyright{acmlicensed}\acmConference[ESEM '20]{ESEM '20: ACM / IEEE International Symposium on Empirical Software Engineering and Measurement (ESEM)}{October 8--9, 2020}{Bari, Italy}
\acmBooktitle{ESEM '20: ACM / IEEE International Symposium on Empirical Software Engineering and Measurement (ESEM) (ESEM '20), October 8--9, 2020, Bari, Italy}
\acmPrice{15.00}
\acmDOI{10.1145/3382494.3410681}
\acmISBN{978-1-4503-7580-1/20/10}

\begin{document}

\title[Adoption and Effects of Software Engineering Best Practices in Machine Learning]{Adoption and Effects of Software Engineering Best Practices in~Machine Learning}


\author{Alex Serban}
\email{a.serban@cs.ru.nl}
\affiliation{%
  \institution{ICIS, Radboud University}
  \institution{Software Improvement Group}
  \city{Amsterdam}
  \state{The Netherlands}
}

\author{Koen van der Blom}
\author{Holger Hoos}
\author{Joost Visser}
\affiliation{%
	\institution{LIACS, Leiden University}
	\city{Leiden}
	\state{The Netherlands}
}


\begin{abstract}
\textbf{Background.}
The increasing reliance on applications with \ac{ML} components calls for mature engineering techniques that ensure these are built in a robust and future-proof manner.
\\
\textbf{Aim.}
We aim to empirically determine the state of the art in how teams develop, deploy and maintain software with \ac{ML} components.
\\
\textbf{Method.}
We mined both academic and grey literature and identified 29 engineering best practices for \ac{ML} applications.
We conducted a survey among \finalFiltered~practitioners to determine the degree of adoption for these practices and to validate their perceived effects.
Using the survey responses, we quantified practice adoption, differentiated along demographic characteristics, such as geography or team size.
%
We also tested correlations and investigated linear and non-linear relationships between practices and their perceived effect using various statistical models.
\\
\textbf{Results.}
Our findings indicate, for example, that larger teams tend to adopt more practices, and that traditional software engineering practices tend to have lower adoption than \ac{ML} specific practices.
Also, the statistical models can accurately predict perceived effects such as agility, software quality and traceability, from the degree of adoption for specific sets of practices.
Combining practice adoption rates with practice importance, as revealed by statistical models, we identify practices that are important but have low adoption,
as well as practices that are widely adopted but are less important for the effects we studied.
\\
\textbf{Conclusion.}
Overall, our survey and the analysis of responses received provide a quantitative basis for
assessment and step-wise improvement of practice adoption by \ac{ML} teams.
\end{abstract}



\maketitle

\acrodef{ML}[ML]{machine learning}
\acrodef{AI}[AI]{artificial intelligence}
\acrodef{RF}[RF]{random forest}
\acrodef{DL}[DL]{deep learning}
\acrodef{NN}[NN]{neural networks}
\acrodef{DNN}[DNN]{deep neural network}
\acrodef{SE}[SE]{software engineering}

\section{Introduction}
\label{sec:intro}

The adoption of \ac{ML} components in production-ready applications demands strong engineering methods to ensure robust development, deployment and maintenance.
While a wide body of academic literature acknowledges these challenges~\cite{lwakatare2019taxonomy, ishikawa2019engineers, wan2019does, arpteg2018software, khomh2018software}, there is little academic literature to guide practitioners.
In fact, a large part  of the  literature concerning engineering practices for \ac{ML} applications can be classified as grey literature~\cite{garousi2019guidelines} and consists of blog articles, presentation slides or white papers. 

In this work, we aim to determine the state of the art in how teams develop, deploy and maintain software solutions that involve \ac{ML} components.
%
Towards this goal, we have first distilled a set of 29 engineering best practices from the academic and grey literature.
These practices can be classified as \emph{traditional} practices, which apply to any software application, \emph{modified} practices, which were adapted from traditional practices to suit the needs of \ac{ML} applications, and completely \emph{new} practices, designed for \ac{ML} applications.

In order to validate the adoption and relevance of the practices we ran a survey among \ac{ML} practitioners, with a focus on teams developing software with \ac{ML} components.
The survey was designed to measure the adoption of practices and also to assess the effects of adopting specific sets of practices.
We obtained \finalFiltered~valid responses and analysed 29 practices and their influence on 4 different effects.

The main contributions of our work are as follows.
Firstly, we summarise academic and grey literature in a collection of best practices.
This body of information can be used by practitioners to improve their development process and serves as a gateway to literature on this topic.
Secondly, we determine the state of the art by measuring the adoption of the practices.
These results are used to rank the practices by adoption level and can serve to assess the popularity of particular practices.
Thirdly, we investigate the relationship between groups of practices and their intended effects, through different lenses~--~by training a linear regression model to check if the intended effect is dependent on the practices and by training more sophisticated regression models, using a variety of ML approaches (including AutoML) to predict the effects from the practices.
Lastly, we investigate the adoption of practices based on the data type being processed and based on the practice categories introduced above (traditional, modified, new).

Our results suggest that the practices apply universally to any \ac{ML} application and are largely independent of the type of data considered.
Moreover, we found a strong dependency between groups of practices and their intended effect.
Using the contribution of each practice to the desired effect (extracted from our predictive models) and their adoption rate, we outline a method for prioritising practice improvements tailored 
for achieving specific effects, such as increased traceability or software quality.
While our study is restricted to \ac{ML}, rather than the broader and less clearly delineated field of \ac{AI}, many of our findings may have wider applications, as we will briefly discuss in Section~\ref{sec:conclusion}.

The remainder of the paper is organised as follows.
We first discuss background and related work (Section~\ref{sec:background}).
Next, we describe the process and results of mining practices from literature (Section~\ref{sec:mining}).
A description of the design of our study (Section~\ref{sec:design}) is followed by a presentation of the survey results regarding the adoption of practices (Section~\ref{sec:results}) and a deeper analysis of the relationship between the practices and their effects (Section~\ref{sec:analysis}). Finally, we discuss interpretation and limitations of our findings (Section~\ref{sec:discussion}) and close with general conclusions and remarks on future work (Section~\ref{sec:conclusion}). Our survey questions, data, and code for analysis and visualisation are publicly available~\cite{supplementary}.

\section{Background and Related Work}
\label{sec:background}

\subsubsection*{\textbf{Engineering challenges posed by \ac{ML}}}
As \ac{ML} components are developed and deployed, several engineering challenges specific to the \ac{ML} software development life-cycle emerge~\cite{lwakatare2019taxonomy, ishikawa2019engineers, wan2019does, arpteg2018software, khomh2018software}.
\citeauthor{arpteg2018software}~\cite{arpteg2018software} identified a set of 12 challenges that target development, deployment and organisational issues. 
In particular, managing and versioning data during development, monitoring and logging data for deployed models and estimating the effort needed to develop \ac{ML} components present striking differences with the development of traditional software components.

Similarly, \citeauthor{ishikawa2019engineers}~\cite{ishikawa2019engineers} as well as \citeauthor{wan2019does}~\cite{wan2019does} have studied how software engineers perceive the challenges related to \ac{ML} and how \ac{ML} changes the traditional software development life-cycle.
Both studies ran user surveys with a majority of respondents from Asia.
We could not find a similar study without this regional bias.
Nonetheless, both publications concluded that testing and ensuring the quality of \ac{ML} components is particularly difficult, because a test oracle is missing, the components often behave nondeterministically, and test coverage is hard to define.
In order to better classify the challenges raised by \ac{ML} components, \citeauthor{lwakatare2019taxonomy} introduced a comprehensive taxonomy~\cite{lwakatare2019taxonomy}.


\subsubsection*{\textbf{White and grey literature analysis.}}
In search for ways to meet the challenges presented earlier, we mined the literature and collected \ac{SE} best practices for \ac{ML}.
We observed that the majority of literature on this topic consists of so called grey literature~\cite{garousi2019guidelines}~--~\ie,~blog articles, presentation slides or white papers from commercial companies~--~while there is relatively little academic literature.
\citeauthor{garousi2019benefitting}~\cite{garousi2019guidelines} showed that, if used properly, grey literature can benefit \ac{SE} research, providing valuable additional information. However, this literature must be used with care, because it does not contain strong empirical evidence to support its claims~\cite{garousi2019benefitting}.
We decided to included the grey literature in our literature search, using the process described by~\citeauthor{garousi2019guidelines}~\cite{garousi2019guidelines}, because: (1) coverage of the subject by academic literature is rather incomplete, (2) contextual information is important for the subject of study~--~\ie,~practitioners may have different opinions than scientists on what qualifies as best practices~--~and (3) grey literature may corroborate scientific outcomes with practical experience.


\subsubsection*{\textbf{Related work}}
We focus on peer-reviewed related work that proposes, collects or validates engineering best practices for \ac{ML}.
One of the initial publications on this topic is the work of \citeauthor{sculley2015hidden}~\cite{sculley2015hidden}, which used the framework of technical debt to explore risk factors for \ac{ML} components.
In particular, they argued that \ac{ML} components have a stronger propensity to incur technical debt, because they have all maintenance problems specific to traditional software as well as a set of additional issues specific to \ac{ML}.
They also presented a set of anti-patterns and practices aimed at avoiding technical debt in systems using \ac{ML} components.
Compared to \cite{sculley2015hidden}, we introduce a broader set of practices, applicable to more effects than technical debt.
Nonetheless, some of their suggestions, which are specific to engineering, are included in our catalogue of practices.

\citeauthor{breck2017ml}~\cite{breck2017ml} introduced 28 tests and monitoring practices that target different stages of the development process for \ac{ML}.
They also proposed a list of benefits resulting from implementing the tests and developed
a model to score test practice adoption, aimed at measuring technical debt. 
Again, the practices dedicated to \ac{SE} from~\cite{breck2017ml} have been included in our catalogue.
On the same topic, \citeauthor{zhang2020machine} introduced a survey on testing techniques for \ac{ML} components~\cite{zhang2020machine}, which~--~in contrast to the broader approach taken in~\cite{breck2017ml}~--~only targets testing \ac{ML} models.

To identify challenges faced by small companies in developing \ac{ML} applications, \citeauthor{de2019understanding} ran interviews with 7 developers~\cite{de2019understanding}.
Afterwards, they proposed and validated a set of checklists to help developers overcome the challenges faced. 
Although the validation session is not thorough (it only included a focus group with 2 participants), some of the items in the checklists qualify as best practice candidates.
Some of these practices are included in our  catalogue and our survey further confirms their relevance and adoption.

\citeauthor{washizaki2019studying}~\cite{washizaki2019studying} studied and classified software architecture design patterns and anti-patterns for \ac{ML}, extracted from white and grey literature. 
Many of these patterns are application and context specific, \ie,~they depend on the architectural style or on the type of data used.
The patterns are of a general character and the ones similar to recommendations we found in literature were included in our catalogue of practices.

\citeauthor{amershi2019software} conducted a study  internally at Microsoft, aimed at collecting challenges and best practices for \ac{SE} used by various teams in the organisation~\cite{amershi2019software}.
They reported on a broad range of challenges and practices used at different stages of the software development life cycle.
Using the experience of the respondents and the set of challenges, they built a maturity model to assess each team.
However, the set of challenges and reported practices are broad and often not actionable.
Moreover, they represent the opinions of team members from Microsoft, where typically more  resources are dedicated to ensuring adoption of best practices than within smaller companies.
In our work, we aim to bridge this gap by running a survey with practitioners with various backgrounds and by presenting a set of actionable, fine-grained best practices.

\section{Mining practices from literature}
\label{sec:mining}

\begin{table}[t]
	\caption{Successful search queries. The table shows the base queries, for which any variant (described in text) led to a valid source and at least one practice.}

	\label{tbl:search}
	\begin{tabular}{ll}
    \toprule
		Query & \specialcell{Documents } \\
    \midrule
		software engineering for machine learning  &  \cite{amershi2019software} \\ 
		data labeling best practices & \cite{DOML, DLB, DCML, DLML} \\ 
		machine learning engineering practices &  \cite{Rs4ML, MLOps, CD4ML} \\ 
		software development machine learning & \cite{SD4DL} \\ 
		machine learning production & \cite{MLPROD, DMP} \\
		machine learning production practices & \cite{BPMLI, MMLP, TFX, BPDL}  \\ 
		machine learning deployment & \cite{TDBML} \\ 
		machine learning deployment practices & \cite{MLArch} \\ 
		machine learning pipelines practices & \cite{MDLOPS} \\ 		
		machine learning operations & \cite{OPML} \\
		machine learning versioning & \cite{VML} \\
		machine learning versioning practices & \cite{PMLPP} \\
    \bottomrule
	\end{tabular}
\end{table}

\subsubsection*{\textbf{Document Search.}}
In addition to the publications discussed in Section~\ref{sec:background}, we searched the academic and grey literature on the topic of \ac{SE} best practices for \ac{ML} applications.
We used both Google and Google Scholar, for which we compiled a common set of queries.
The keywords used for querying suggest different steps in the development cycle, \eg,~development, deployment, operations, \etc~For each query, we also developed two variants, by (1) replacing the term `machine learning' with `deep learning' whenever possible, and (2) removing stop words and composing a Boolean AND query from the remaining key words.
As an example of the second variant, consider the query ``software engineering'' AND ``machine learning'', stemming from the query ``software engineering for machine learning''.
All queries were submitted to Google and Google Scholar, and the first 5 result pages were manually inspected.

A total of 64 queries, including variants, were used, and 43 of the resulting articles were selected for initial inspection.
In order to avoid search engine personalisation, all queries were sent
from a public network, with an extension that removes browser cookies.

\subsubsection*{\textbf{Document classification.}}
Based on criteria formulated in~\cite{garousi2019guidelines}, such as authoritativeness of the outlet and author as well as objectivity of the style and content, we excluded low-quality documents and classified the remaining documents as either academic literature or grey literature.
Moreover, we filtered for duplicates, because chunks of information were sometimes reused in grey literature.

After classifying and filtering the results, we identified 21 relevant documents, including scientific articles, white papers, blogs and presentation slides, that -- along with the publications introduced in Section~\ref{sec:background} -- were used to mine \ac{SE} best practices for \ac{ML}.
Other relevant sources were selected through a snowball strategy, by following references and pointers from the initial articles.

Table~\ref{tbl:search} lists the successful search terms (without variants), from which at least one document passed the final selection.
Whenever the queries had common results, we only considered relevant the first query.
The second column in Table~\ref{tbl:search} shows the
documents selected from the base queries and their variants.

\subsubsection*{\textbf{Extracting a common taxonomy for the practices.}}
Many of the selected documents provide, or implicitly presume, a grouping of practices
based on development activities specific to \ac{ML}.
For example, \citeauthor{amershi2019software}~\cite{amershi2019software} present a nine-stage \ac{ML} pipeline.
Alternatively, \citeauthor{CD4ML}~\cite{CD4ML} partition similar activities into six pipeline stages.
All processes have roots in early models for data mining, such as CRISP-DM~\cite{wirth2000crisp}.

While no single partitioning of \ac{ML} activities emerged as most authoritative, we were able to reconstruct a broad taxonomy that is compatible with all partitionings found in the literature.
We will use this categorisation to group ML development practices and to structure our survey and subsequent discussion of our findings:

\begin{itemize}
	\item Data~-~Practices that come before training, including collecting and preparing data for training \ac{ML} models.
	\item Training~-~Practices related to planning, developing and running training experiments.
	\item Deployment~-~Practices related to preparing a model for deployment, deploying, monitoring and maintaining an \ac{ML} model in production.
	\item Coding~-~Practices for writing, testing, and deploying code. 
	\item Team~-~Practices related to  communication and alignment in a software development team.
	\item Governance~-~Practices that relate to ensuring responsible use of \ac{ML}, including accountability regarding privacy, transparency, and usage of  human, financial, or energy resources.
\end{itemize}

\newcommand{\data}{Data}
\newcommand{\experiment}{Training}
\newcommand{\code}{Coding}
\newcommand{\deploy}{Deployment}
\newcommand{\reference}{References}
\newcommand{\team}{Team}
\newcommand{\ranking}{Rank}
\newcommand{\governance}{Governance}
\newcommand{\type}{Type}
\newcommand{\newpractice}{N}
\newcommand{\modified}{M}
\newcommand{\old}{T}

\num\def\Q38{1}
\num\def\Q37{2}
\num\def\Q48{3}
\num\def\Q44{4}
\num\def\Q34{5}
\num\def\Q43{6}
\num\def\Q47{7}
\num\def\Q82{8}
\num\def\Q65{9}
\num\def\Q66{10}
\num\def\Q35{11}
\num\def\Q60{12}
\num\def\Q62{13}
\num\def\Q36{14}
\num\def\Q56{15}
\num\def\Q55{16}
\num\def\Q61{17}
\num\def\Q33{18}
\num\def\Q63{19}
\num\def\Q42{20}
\num\def\Q64{21}
\num\def\Q32{22}
\num\def\Q40{23}
\num\def\Q57{24}
\num\def\Q59{25}
\num\def\Q80{26}
\num\def\Q53{27}
\num\def\Q41{28}
\num\def\Q39{29}

\begin{table*}[ht]
	\caption{\ac{SE} best practices for \ac{ML}, grouped into 6 classes, together with the practice type, literature references and adoption ranks, where \newpractice~--~new practice, \old~--~traditional practice, \modified~--~modified practice.}
	
	\label{tbl:practices}
	\begin{tabular}{lp{32.4em}llp{6.5em}l}
    \toprule
		Nr. &  Title & Class  & \type &  \reference &  \ranking \\
    \midrule
		1 & \specialcell{Use Sanity Checks for All External Data Sources} & \data & \newpractice  & \cite{DMP,MLOps} & \Q32 \\ 
		2 & \specialcell{Check that Input Data is Complete, Balanced and Well Distributed} & \data &\newpractice & \cite{CTPML, DMP, sculley2015hidden, MMLP, breck2017ml} & \Q33 \\ 
		3 & \specialcell{Write Reusable Scripts for Data Cleaning and Merging} &  \data &\newpractice & \cite{BPMLI, DMP, MLOps} & \Q34 \\ 
		4 & \specialcell{Ensure Data Labelling is Performed in a Strictly Controlled Process}  & \data &\newpractice & \cite{DCML, DLB, DLML, DOML} & \Q35 \\ 
		5 & \specialcell{Make Data Sets Available on Shared Infrastructure (private or public)} & \data &\newpractice & \cite{MMLP, khomh2018software, PMLPP, SD4DL} & \Q36 \\
    \midrule
		6 & \specialcell{Share a Clearly Defined Training Objective within the Team} & \experiment &\newpractice & \cite{MLTEAM, MMLP, Rs4ML} & \Q37 \\ 
		7 &  Capture the Training Objective in a Metric that is Easy to Measure and Understand & \experiment &\newpractice & \cite{MLTEAM, OPML, Rs4ML, DSTEAM} & \Q38 \\ 
		8 & Test all Feature Extraction Code & \experiment & \modified &  \cite{CD4ML, breck2017ml} & \Q40 \\
		9 & Assign an Owner to Each Feature and Document its Rationale & \experiment & \modified & \cite{Rs4ML} & \Q39 \\
		10 & Actively Remove or Archive Features That are Not Used & \experiment &\newpractice & \cite{sculley2015hidden, Rs4ML} & \Q41 \\
		11 & Peer Review Training Scripts & \experiment & \modified & \cite{MLTS} & \Q42 \\ 
		12 & Enable Parallel Training Experiments & \experiment &\newpractice & \cite{CD4ML, MLPROD} & \Q43 \\ 
		13 & Automate Hyper-Parameter Optimisation and Model Selection & \experiment &\newpractice & \cite{automl_book,CSAML} & \Q80 \\ 
		14 & Continuously Measure Model Quality and Performance & \experiment &\newpractice & \cite{Rs4ML, TDBML} & \Q44 \\ 
		15 & Share Status and Outcomes of Experiments Within the Team & \experiment &\newpractice & \cite{BPDL, PMLPP}  & \Q47 \\ 
		16 & \specialcell{Use Versioning for Data, Model, Configurations and Training Scripts} & \experiment & \modified & \cite{BPDL, MDLOPS, MLPROD, MMLP, PMLPP, VML,washizaki2019studying} & \Q48 \\
    \midrule		
		17 & Run Automated Regression Tests & \code &\old & \cite{MLPROD,breck2017ml} & \Q53 \\ 
		18 & Use Continuous Integration & \code &\old & \cite{CD4ML, breck2017ml} & \Q55 \\ 
		19 & Use Static Analysis to Check Code Quality & \code &\old & \cite{visser2016building} & \Q57 \\
    \midrule		
		20 &  Automate Model Deployment &  \deploy & \modified & \cite{MLArch, MLLG, OPML, VML} & \Q56 \\
		21 & Continuously Monitor the Behaviour of Deployed Models & \deploy & \newpractice & \cite{CD4ML, MLLG, MLPROD, TDBML, TFX} & \Q60 \\ 
		22 & Enable Shadow Deployment & \deploy & \modified & \cite{MLLG, TFX, VML,washizaki2019studying} & \Q59 \\ 
		23 & Perform Checks to Detect Skews between Models & \deploy & \newpractice & \cite{CD4ML, Rs4ML, TDBML, TFX} & \Q61 \\ 
		24 & Enable Automatic Roll Backs for Production Models &  \deploy & \modified & \cite{CD4ML, MLLG} & \Q62 \\ 
		25 & \specialcell{Log Production Predictions with the Model's Version and Input Data} &  \deploy & \modified & \cite{MDLOPS, MLGov, MMLP} & \Q63 \\
    \midrule		
		26 & Use A Collaborative Development Platform & \team &\old & \cite{booch2003collaborative, storey2016social} & \Q82 \\
		27 & Work Against a Shared Backlog & \team  &\old & \cite{sedano2019product, sutherland2013scrum} & \Q65 \\ 
		28 & Communicate, Align, and Collaborate With Multidisciplinary  Team Members &  \team &\old & \cite{faraj2000coordinating} & \Q66 \\
    \midrule		
		29 &  Enforce Fairness and Privacy  & \governance & \newpractice & \cite{MLFAIR, MLRES,breck2017ml} & \Q64 \\
    \bottomrule
    \end{tabular}
\end{table*}

\subsubsection*{\textbf{Compiling  a catalogue of practices.}} From the selected documents we compiled an initial set of practices using the following methodology.
First, we identified all practices, tests or recommendations that had similar goals. 
In some articles, the recommendations only suggest the final goal~--~\eg,~ensure that trained \ac{ML} models can be traced back to the data and training scripts used~--~without providing details on the steps needed to achieve it.
In other publications, the recommendations provided detailed steps used to achieve the goals~--~\eg,~use versioning for data, models, configurations and training scripts~\cite{VML, MLPROD, MMLP}.
In this example, traceability is an outcome of correctly versioning all artefacts used in training.
Whenever we encountered similar scenarios, we selected or abstracted actionable practices and added the high-level goals to a special group, which we call ``Effects'' and describe in  Table~\ref{tbl:practices_effects}.

Next, we assessed the resulting practices and selected
those
specifically related to engineering or to the organisation of engineering processes.
This initial selection gave us
23 practices, which naturally fall into 4 out of the 6 classes introduced above.
While this set of practices reflected the \ac{ML} development process, it lacked practices from traditional \ac{SE}.
Given that practitioners with a strong background in \ac{ML} might be unaware of the developments in \ac{SE}, in a third stage, we complemented the initial set of practices with 6 practices from traditional \ac{SE}~--~three of a strictly technical nature, falling into the ``Coding'' class, and three relating to social aspects, falling into the ``Team'' class.
We selected these practices because we consider them challenging, yet essential in software development.

The resulting 29 practices are listed in Table~\ref{tbl:practices} and the effects in Table~\ref{tbl:practices_effects}.
The practices are available to practitioners in a more elaborate format in an online catalogue\footnote{\url{https://se-ml.github.io/practices/}}, consisting of detailed descriptions and concise statements of intent, motivation, related practices, references and an indication of difficulty.
A curated reading list with these references, further relevant literature as well as a selection of supporting tools is maintained online\footnote{\url{https://github.com/SE-ML/awesome-seml}}.
%
%
\section{Study design}
\label{sec:design}
We validated the set of practices with both researchers and practitioners through a survey.
For this, we designed a descriptive questionnaire asking respondents if the \emph{team} they are part of adopts, in their \ac{ML} projects, the practices we identified earlier.
Before distributing the survey, we interviewed five practitioners with diverse backgrounds, in order to check if any information from the survey was redundant or whether important practices were missing.

\subsubsection*{\textbf{Questionnaire}}
In designing the questionnaire used in our survey, we followed the recommendations 
of Kitchenham et al.~\cite{kitchenham2002principles2} 
and \citeauthor{ciolkowski2003practical}~\cite{ciolkowski2003practical}.
We designed a cross-sectional observational study~\cite{kitchenham2002principles2},~\ie,~participants were asked at the moment of filling the questionnaire if they adopted the recommended practices.
Several preliminary questions were designed to specifically assign participants to groups. %
This renders the study a concurrent control study, in which participants are not randomly assigned to groups~\cite{kitchenham2002principles2}.

The target audience were \emph{teams} of practitioners using \ac{ML} components in a project.
Specific preliminary questions were added to allow filtering between teams that build and deploy \ac{ML} applications, use \ac{ML} and do not build an application or do not use \ac{ML} at all.
We consider that a significant amount of engineering is also needed in research (where \ac{ML} may be used without building deployed applications), especially in running large-scale deep learning experiments, and would like to verify which practices are relevant in this context.
Team profile (\eg,~tech company, governmental organisation), team size (\eg,~1 person, 6-9 persons), team experience (\eg,~most of us have 1-2 years of experience), 
and the types of data used (\eg,~tabular data, images) were also included in the preliminaries.
In total, the preliminaries contained 5 questions that were later used to group participants and filter out noise.

Then, 31 questions followed, with standard answers, mapped onto the practices from Table~\ref{tbl:practices}.
In two cases, multiple questions map onto the same practice; for example, continuous integration is achieved by automating the build process and running it at each commit. 
In the questionnaire, we asked two questions, one for each action, although we compiled the answers to one best practice.

We used standard answers, on a Likert scale with four possible answers, in order to avoid the middle null-point strategy of answering~\cite{hofmans2007bias}.
The labels were chosen in order to reflect the degree of adoption, rather than the level of agreement~\cite{rohrmann2007verbal}.
This allowed the practices to be expressed impartially~--~\eg~,~``our software build process is fully automated''~--~and the answers to express degrees of adoption~--~\eg,~``not at all'' or ``completely''~--~instead of degrees of agreement such as ``agree'' or ``strongly agree''.
This strategy eliminated confusing questions and answers, which may lead to an extreme null-point bias~\cite{hofmans2007bias}.
Whenever the answer scale did not match the full range of answers, we added specific answers which helped to avoid noisy results;
for example, in the questions about data labelling, we added the answer ``we do not use data labels'', which accounts for unsupervised learning scenarios.

The questionnaire ended with a section on the perceived effects of adopting the practices.
This enabled us to test the hypothesis that adopting a group of practices helps to achieve an effect.
The four questions on perceived effects are shown in Table~\ref{tbl:practices_effects}.

Although the questionnaire has 45 questions, we employ optimisation techniques, such as automatically moving to the next question once an answer is selected, to reduce the time required for completing our questionnaire to \sTime~minutes on average.

\subsubsection*{\textbf{Pilot interviews}}

\begin{table}[t]
	\caption{Profiles of the pilot interview subjects.}
	\label{tbl:interviews}
	\begin{tabular}{l l l l }
    \toprule
		Id & Company Profile & Team Size & Experience  \\
    \midrule
		P1 & Tech Startup & 5-6 ppl. & 1-2 years  \\  
		P2 & Tech company & 10-15 ppl. &  >5 years \\ 
		P3 & Research lab & 5-6 ppl. & 2-5 years  \\ 
		P4 & Tech Startup & 10-15 ppl. & 2-5 years  \\ 
		P5 & Non-tech company & 6-9 ppl. & 1-2 years  \\ 
    \bottomrule
	\end{tabular}
\end{table}

Before distributing the survey  broadly, we invited five practitioners with distinct backgrounds~--~ranging from tech startups to large tech companies~--~to an interview. We asked them to answer a set of questions regarding the challenges they face and the most important engineering practices they adopt.
All interviewees were also asked to fill out and comment on the questionnaire.
Since the survey is focused on teams, in Table~\ref{tbl:interviews} we present the team and company profiles for each interviewee;
all interviewees use \ac{ML} for a project. 
Moreover, P4 is part of a team that builds platforms to support the \ac{ML} process and uses distinct \ac{ML} projects to test the platforms.

The biggest challenges faced by the interviewees were: ensuring data quality and data documentation (P5), data versioning and freshness (P2),  scalability (P1, P4) and communication with other departments inside the company (P5).
For each challenge mentioned, there is at least one practice addressing it in Table~\ref{tbl:practices}.
The most important engineering practices mentioned were: using version control for data and models (P1, P4), continuous deployment (P2, P5) and model maintenance (P2).
Several practices to address these challenges were already listed in Table~\ref{tbl:practices}.

After completing the questionnaire, all interviewees agreed with the relevance of all the practices  we 
listed and did not consider any of them redundant. 
The interviewees suggested that some questions needed additional allowable answers, 
to cover the range of possible responses and to avoid bias.
For example, for a question about the labelling process, we added 
``we do not use labels'' to avoid forcing users of unsupervised learning to resort to ``not at all''.

We used the feedback from the interviews to refine the questionnaire, adding answers to four questions and rewording others.

\subsubsection*{\textbf{Distribution}}
After the pilot interviews, 
our survey was distributed using a snowball strategy.
Initially, we reached out to our network of contacts and to the authors of the publications used to extract the practices, asking them to distribute the survey through their networks.
Moreover, we openly advertised the survey through channels commonly used by practitioners, such as Twitter, Medium, HackerNoon, Dev.to and the Meetup groups for \ac{ML} in several cities.

\section{Findings on Practice Adoption}
\label{sec:results}

\newcommand{\graphWidth}{4.2cm}
 \begin{figure*}[t]
 	\centering
    \hspace*{-5.5ex}
 	\subfloat[Respondents grouped by \newline regions.\label{fig:continents}]{\includegraphics[width=3.8cm, keepaspectratio,valign=t]{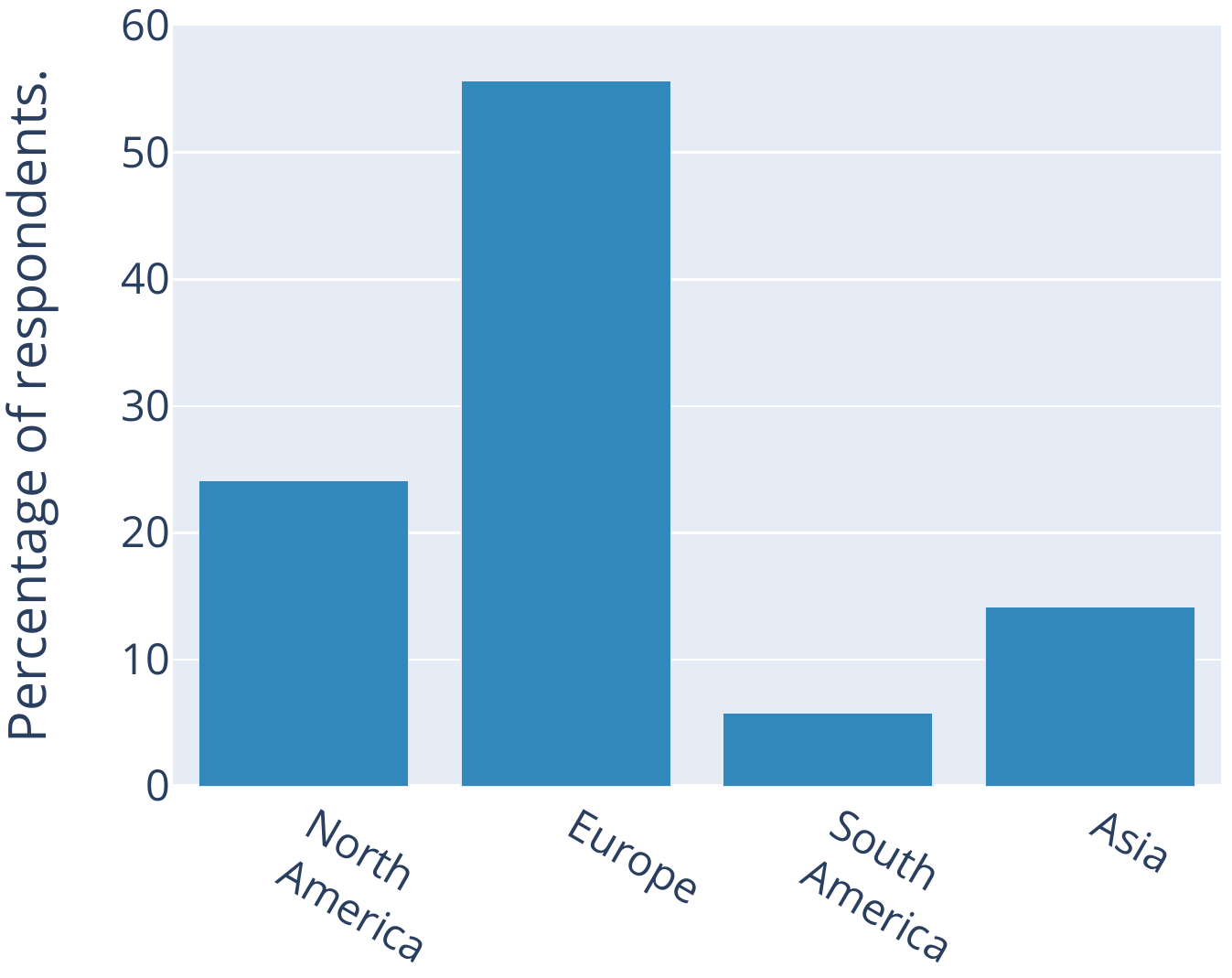}}
\hspace*{2ex}
 	\subfloat[Respondents grouped by \newline organisation type.\label{fig:org_type}]{\includegraphics[width=3.8cm, keepaspectratio, valign=t]{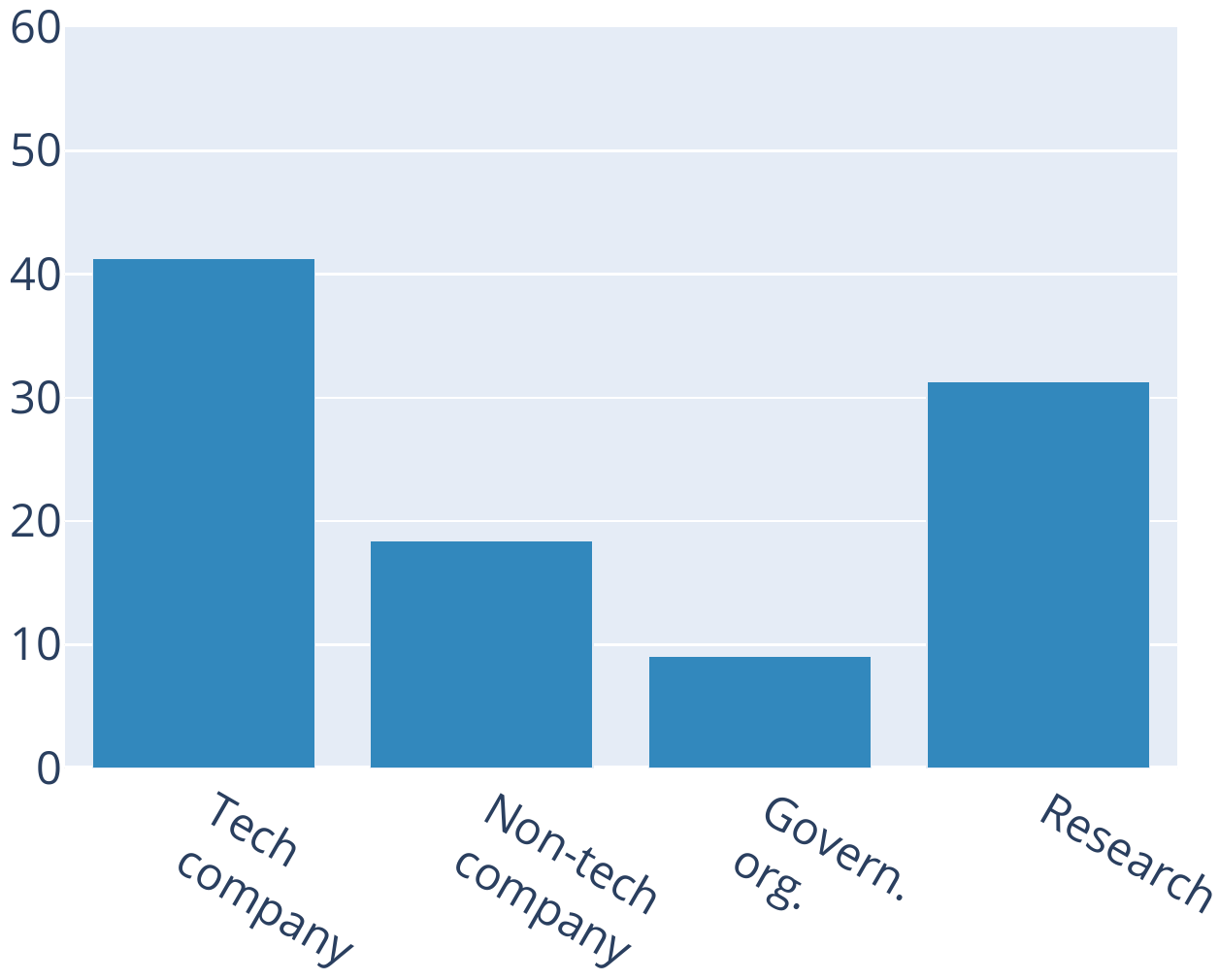}}
 \hspace*{3ex}
 	\subfloat[Respondents grouped by \newline team size.\label{fig:team_size}]{\includegraphics[width=3.8cm, keepaspectratio, valign=t]{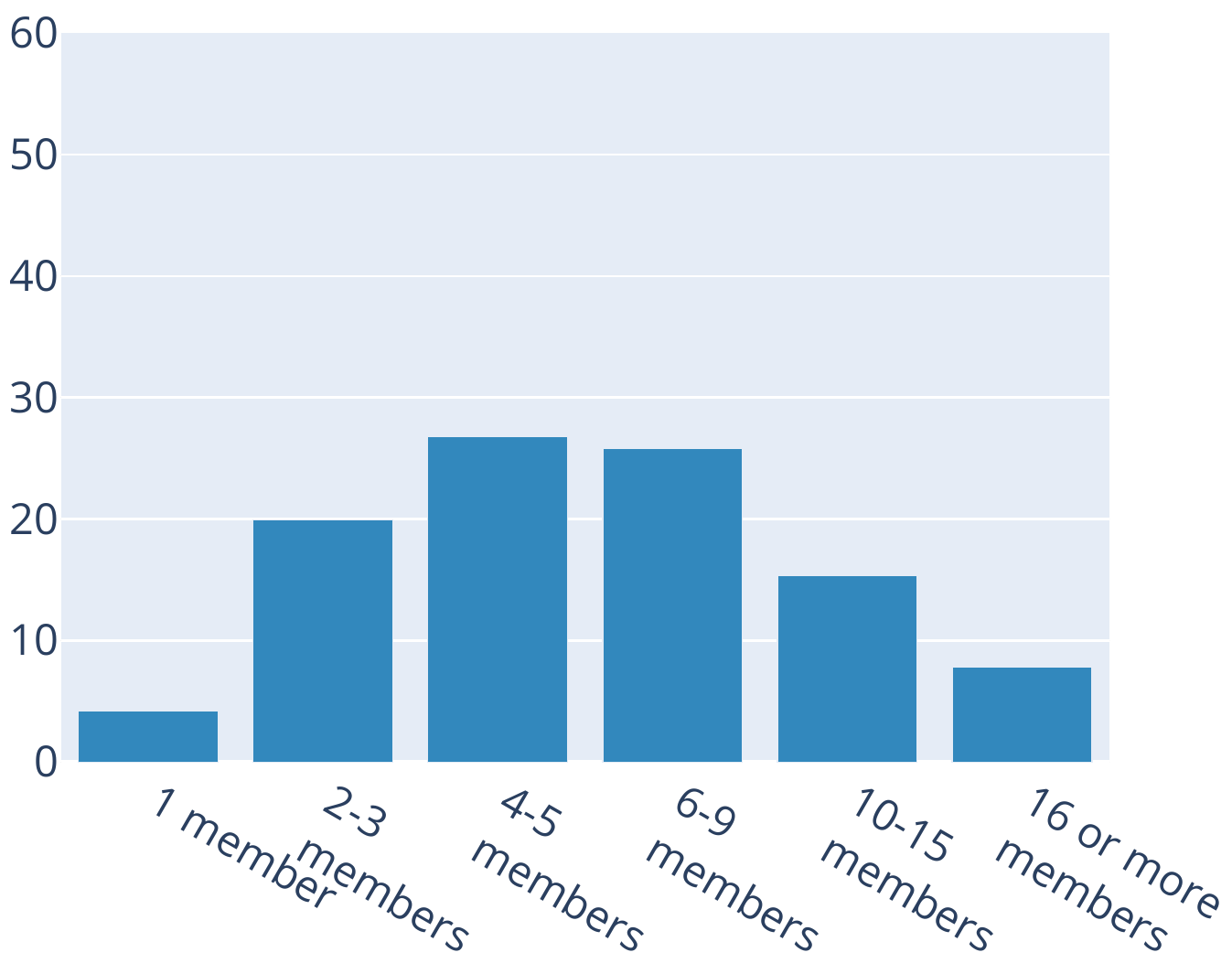}}
\hspace*{2ex}
 	\subfloat[Respondents grouped by \newline team experience.\label{fig:team_experience}]{\includegraphics[width=3.8cm, keepaspectratio,    valign=t]{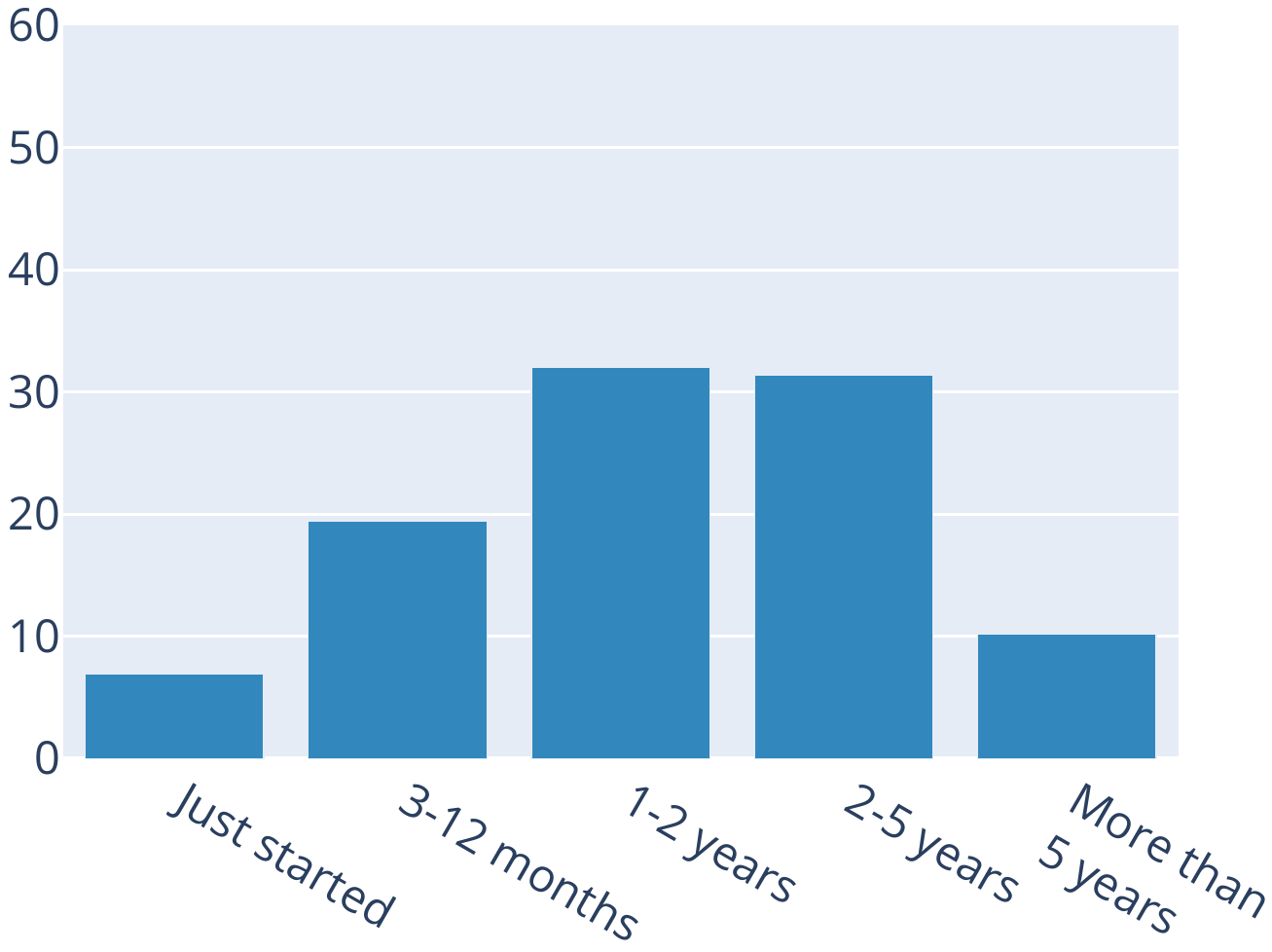}}
 \hspace*{5ex}
 	\caption{Demographic information describing the survey participants. All plots show the percentage of respondents, grouped by various demographic factors.}
   	\label{fig:demographics}
 \end{figure*}
 \begin{figure*}[ht]
 	\centering
 	\subfloat[Adoption of practices \newline grouped by regions.\label{fig:adoption_continents}]{\includegraphics[width=3.9cm, keepaspectratio,valign=t]{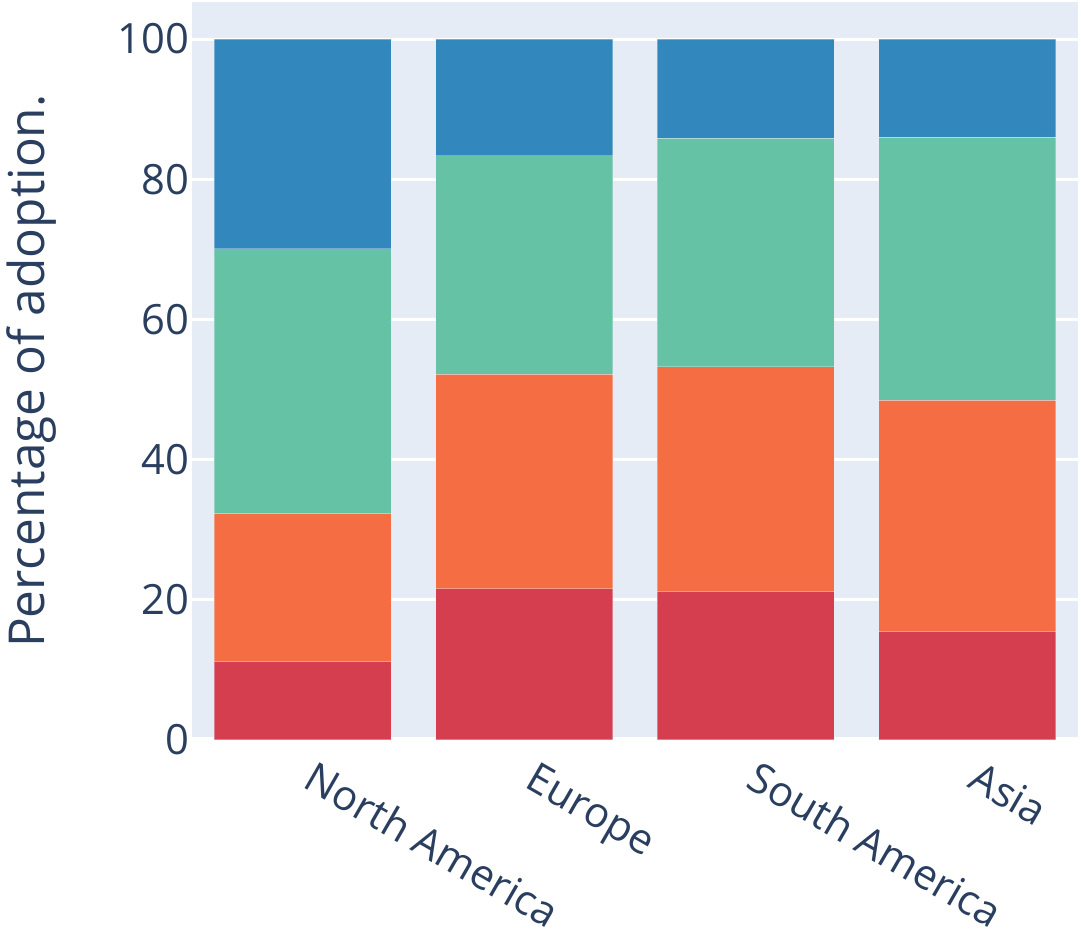}}
 	\quad
 	\subfloat[Adoption of practices \newline grouped by organisation type.\label{fig:adoption_orgtype}]{\includegraphics[width=3.62cm, keepaspectratio,valign=t]{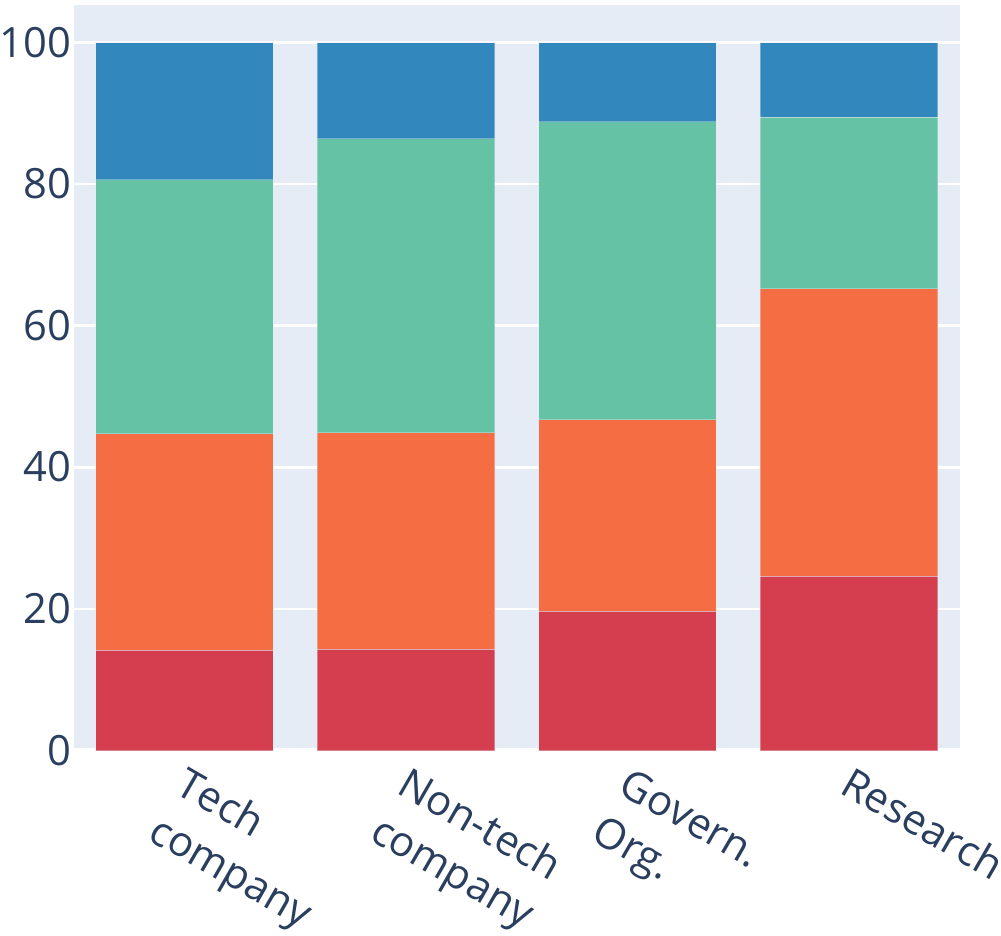}}
 	\qquad
 	\subfloat[Adoption of practices \newline grouped by team size.\label{fig:adoption_teamsize}]{\includegraphics[width=3.85cm, keepaspectratio,valign=t]{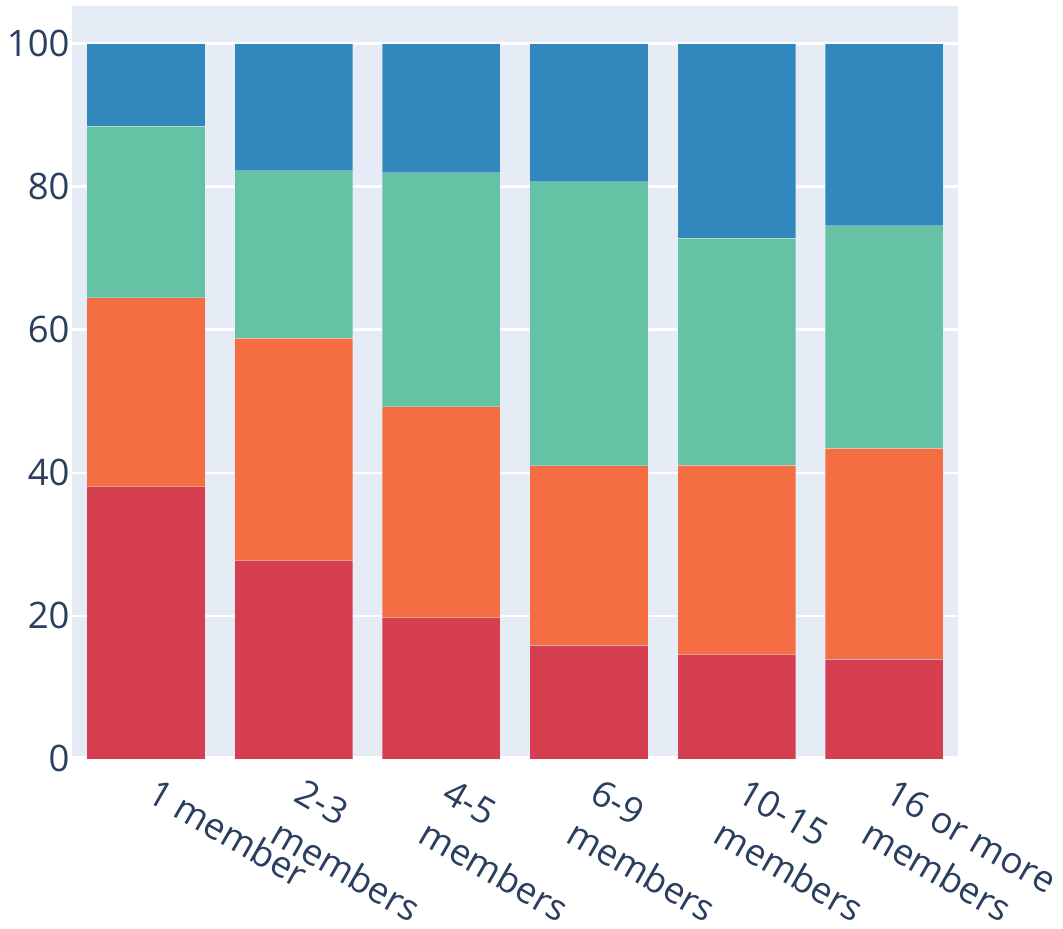}}
 	\quad
 	\subfloat[Adoption of practices \newline grouped by team experience.\label{fig:adoption_teamexperience}]{\includegraphics[width=5cm, height=3.4cm,valign=t]{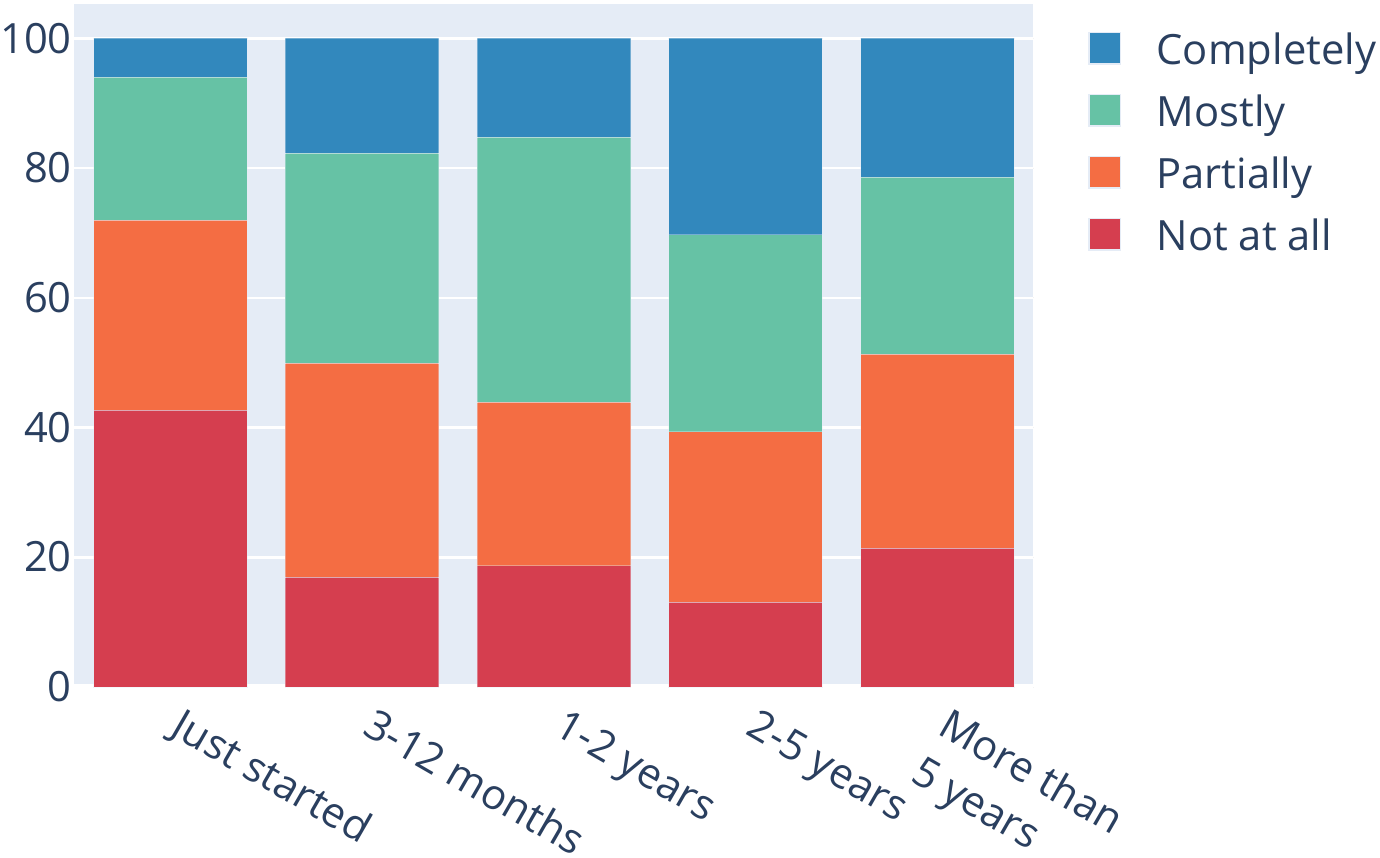}}
 	\caption{Adoption of practices grouped by various demographic factors. All plots show the percentage of answers, grouped by the answer types illustrated in the plot legend.}
  	\label{fig:adoption_demographics}
 \end{figure*}
In total we obtained \allAnswers~valid responses to our survey, after filtering out incomplete answers or respondents that spent too little time
to have given serious answers (under 2 minutes).
From this initial set, we discarded \notInTeam~answers from respondents who were not part of a team using \ac{ML}.
Moreover, we applied fine-grained filtering, using the percentage of questions that were answered in the prerequisites (at least \percPrereq~\%) and in the practice adoption questions (at least \percPractice~\%), resulting in \emph{\finalFiltered~complete responses}.
Whenever not mentioned otherwise, the analysis will be based on these responses.

\subsubsection*{\textbf{Demographics.}}
\label{subsec:demographics}
Using the initial preliminary questions, we provide a demographic characterisation of the respondents in Figure~\ref{fig:demographics}.
Firstly, we grouped the answers using the location attributes and present the results in Figure~\ref{fig:continents}.
We observe that Europe has an overall higher contribution, although other regions are also well represented.
This possible bias will be discussed later in this section, when analysing the answers for each region.

Figure~\ref{fig:org_type} illustrates the percentage of respondents grouped by the organisation type.
The higher percentages are for teams working in tech companies (\eg~social media platforms, semiconductors) and research labs.
These results are not surprising, since both research and adoption of \ac{ML} is driven by these two classes of practitioners.
Nonetheless, non-tech companies (\eg~enterprises, banks) and governmental organisations are also well represented.

In the last two plots we show the percentage of answers grouped by team size~--~Figure~\ref{fig:team_size}~--~and team experience~--~Figure~\ref{fig:team_experience}.
We observe that most teams have between 4-5 and 6-9 members, corresponding to normal software development teams (as recommended, for example, in Agile development).
Similarly, most teams have between 1-2 years and 2-5 years of experience, which is an anticipated result, since these intervals correspond to the recent growth in popularity of \ac{ML} among practitioners.
Overall, the demographic information indicates that our data is well balanced and diverse.

Next, we analysed the adoption of practices grouped by the demographic factors introduced earlier.
We display the answers from the practice questions in Figure~\ref{fig:adoption_demographics}, grouped and normalised using the Likert scale used in the survey.
Figure~\ref{fig:adoption_continents} shows the percentage of answers grouped by regions.
As discussed earlier, Europe is somewhat over-represented in our data set.
However, the adoption of practices for Europe does not present striking differences when compared to South America or Asia.
Conversely, the respondents from North America have a significant higher number of adopted practices (corresponding to answers ``Completely'' or ``Mostly'') than other regions.
Since this region is well represented in our set of responses, it is likely that practitioners from North America have a higher adoption of practices.
Moreover, since Europe does not present striking differences with other regions, it is likely that little bias is introduced by its over-representation.

Figure~\ref{fig:adoption_orgtype} shows the adoption of practices grouped by the organisation type.
We observe that tech companies have a higher rate of complete adoption than others.
Research organisations tend to have lower practice adoption.
This could reflect that they are aware of the practices, but only develop prototypes, for which adoption is not needed, or partial adoption is sufficient.
In fact, for non-deployment practices only, adoption rates are similar.

For team size~--~Figure~\ref{fig:adoption_teamsize}~--~we observe a trend towards higher adoption of practices (and also lower percentage of practices that were not adopted at all) as team size increases.
This could be caused by better task distribution among team members, or it could be a result of larger teams including members with different backgrounds.
Similarly, for team experience, there is a trend towards higher adoption of practices as the team experience increases, as seen in Figure~\ref{fig:adoption_teamexperience}.
These results were anticipated, since more experience or a deeper involvement in technology exposes team members to the knowledge needed to adopt best practices.
A contrasting trend can be observed only for teams with more than 5 years of experience, where the percentage of practices that are only partially or not at all adopted increases slightly.
This result may reveal that practitioners who started very early may be unaware of practices that are developed recently.

These results confirm our questions were clear and achieved their goals, and that the answer scale did not introduce bias.

\begin{table}[t]
    \caption{Adoption of practices based on the practice type.}
    \label{tbl:adoption_ranks}
    \centering
    \begin{tabular}{lccc}
        \toprule
         \specialcell{Practice \\ Type} & \specialcell{At least \\ high adoption} & \specialcell{At least \\ medium adoption} & \specialcell{At least \\ low adoption}  \\
        \midrule
         Traditional & 15.6\% & 47.8\% & 76.8\% \\
         Modified & 11.3\% & 42.0\% & 76.9\% \\
         New & \textbf{16.9}\% & \textbf{50.0}\% & \textbf{83.9\%} \\
        \bottomrule
    \end{tabular}
\end{table}

\subsubsection*{\textbf{Practice adoption ranking.}}
\label{subsec:adoption}

We now explore the adoption of practices, based on the practice types discussed in Section~\ref{sec:intro}.
In particular, we are interested in finding out whether traditional \ac{SE} practices are equally adopted in \ac{ML} engineering and which new or modified practices are popular among practitioners.
Moreover, we also comment on the least and most adopted practices.

The practices are classified as follows: (1) \emph{new} practices, designed specifically for the \ac{ML} process, (2) \emph{modified} practices, derived from traditional \ac{SE} practices, but adapted for the \ac{ML} process and (3) \emph{traditional} practices, applied equally in traditional \ac{SE} and \ac{ML}.
This classification is illustrated in the ``Type'' column of Table~\ref{tbl:practices}.

In order to measure the adoption rate of the practices, we devised a ranking algorithm with the following steps:
\begin{enumerate}
\item Compute for each practice the percentage $h$ of respondents with \emph{at least high} adoption (counting ``Completely'' answers), the percentage $m$
with \emph{at least medium} adoption (counting ``Completely'' and ``Mostly''), and the percentage $l$
with \emph{at least low} adoption (counting ``Completely'',  ``Mostly'', and ``Partially''). As an example, for practice 1 we obtained $h=9.62\%$, $m=34.04\%$, and $l=65.92\%$.
\item Convert each percentage into a rank number. For practice 1, we obtained $r_h=22$, $r_m=23$, and $r_l=19$.
\item Take the average of the three ranks for each practice and then rank the practices according to this average. For practice 1, rank 22 was obtained, as can be seen in Table~\ref{tbl:practices}.
\end{enumerate}
%
Thus, the final rank is the average of: the practice rank on \emph{at least high adoption}, its rank on \emph{at least medium adoption}, and its rank on \emph{at least low adoption}.
By accumulating the answers in step 1, we expect to cancel out the noise stemming from fuzzy boundaries between subsequent answer types. 

The results are presented in the ``\ranking''~column of Table~\ref{tbl:practices}, where the highest rank corresponds to the most adopted practice.
We observe that the most adopted practices (practices 6, 7) are related to establishing and communicating clear objectives and metrics for training.
Versioning (practice 16), continuous monitoring of the model during experimentation (practice 14) and writing reusable scripts for data management (practice 3) complete the top 5 positions.
It is interesting to note that the most adopted practices are either new or modified practices, and not traditional \ac{SE} practices.

At the other end of the spectrum, we observe that the least adopted practices (practices 9, 10) are related to feature management.
Writing tests (practice 17), automated hyper-parameter optimisation (practice 13) and shadow deployment (practice 22) complete the 5 least adopted practices.
In general, the least adopted practices require more effort and knowledge.
Some practices, related to testing (practices 8, 17) or documentation (practice 9) are also known to raise issues in traditional \ac{SE}.
Moreover, shadow deployment (practice 22) and automated hyper-parameter optimisation (practice 13) require advanced infrastructure.

\begin{table}[t]
    \caption{Adoption of practices based on the data type.}
    \label{tbl:adoption_data}
    \centering
    \begin{tabular}{lcccc}
        \toprule
        & & & Adoption & \\
        \cline{3-5}
         \specialcell{Data Type} & \specialcell{Perc. of \\[-2pt] respondents}  & \specialcell{At least \\[-2pt] high} & \specialcell{At least \\[-2pt] medium} & \specialcell{At least \\[-2pt] low}  \\
        \midrule
         
         \specialcell{Tabular  Data} & 31.7\% & 18.0\% & 50.1\% & 70.2\% \\
         
         Text & 29.7\% &  19.3\% & 52.6\% & 71.4\% \\
         
         \specialcell{Images, Videos} & 26.4\% & 19.3\% & 50.5\% & 71.5\% \\
         
         \midrule
         
         Audio & 8.8\% & 24.42\% & 55.8\% & 72.6\% \\
         
         \specialcell{Time  Series} & 2.6\% & 28.2\% & 60.3\% & 72.6\% \\
        
          Graphs & 0.5\% & - & - & - \\
        \bottomrule
    \end{tabular}
\end{table}

\begin{table*}[t]
    \caption{Linear regression models describing the dependence of effects on the practices that were initially hypothesised to influence them. For each effect, we report the p-value from the F-test for regression and the $R^2$ coefficient of determination.}
    \label{tbl:practices_effects}
    \centering
    \begin{tabular}{lp{23em}ccc}
        \toprule
         Effects & Description & Practices & p-value & $R^2$  \\
        \midrule
        Agility & The team can quickly experiment with new data and algorithms, and quickly assess and deploy new models & 12, 18, 22, 24, 28  & $7\cdot10^{-4}$ & 0.84 \\
        
        Software Quality & The software produced is of high quality (technical and functional) &  9, 10, 11, 17, 18, 19 &  $5\cdot10^{-3}$  & 0.95 \\
        
        Team Effectiveness & Experts with different skill sets (e.g.,\ data science, software development, operations) collaborate efficiently & 6, 26, 27, 28 &  $1\cdot10^{-5}$ & 0.98 \\
        
        Traceability & Outcomes of production models can easily be traced back to model configuration and input data & 3, 5, 16, 25, 27 & $4\cdot10^{-6}$ &  0.75 \\
        \bottomrule
    \end{tabular}
\end{table*}
\begin{table*}[t]
    \caption{Mean squared error (MSE), $R^2$ and Spearman correlation ($\rho$) between the predicted and the true outcomes for distinct models trained to predict the effects from the practices in the second column, where RF is Random Forest Regression. The results are extracted from a test data set consisting of 25\% of the data.}
    \centering
    \label{tbl:ml_regression}
    \begin{tabular}{lccccc}
        \toprule
         Effects & Practices & \specialcell{MSE / $R^{2}$ / $\rho$ \\ Linear  Regression} & \specialcell{MSE / $R^{2}$ / $\rho$ \\ RF} & \specialcell{MSE / $R^{2}$ / $\rho$ \\ RF Grid Search} & \specialcell{
         MSE / $R^{2}$ / $\rho$ \\ AutoML}  \\
        \midrule
      
        Agility & 12, 18, 20, 21, 22, 28  & 0.69 / 0.44 / 0.68 & 0.27 / 0.78  / 0.92 &  0.25 / 0.80 / 0.92 & \textbf{0.24 / 0.82 / 0.92} \\
      
        Software Quality & 9, 10, 11, 17, 18, 19 &  0.35 / 0.71 / 0.83 & \textbf{0.12 / 0.90 / 0.91} & 0.17 / 0.87 / 0.91 & 0.17 / 0.87 / 0.91  \\
      
        Team Effectiveness & 6, 26, 27, 28 &  0.45 / 0.63 / 0.87 & 0.25 / 0.80 / 0.90 & 0.19 / 0.84 / 0.92 & \textbf{0.18 / 0.85 / 0.92} \\
        
        Traceability & 3, 5, 16, 21, 25, 27 &  0.38 / 0.69 / 0.80  &  0.22 / 0.82 / 0.90  & \textbf{0.21 / 0.83 / 0.93} & 0.22 / 0.82 / 0.93  \\        

        \bottomrule
    \end{tabular}
\end{table*}

In order to compare the adoption of practices grouped by their type, we averaged the three percentages
 described earlier (without transforming them into ranks), for each practice type.
The results are presented in Table~\ref{tbl:adoption_ranks}.
We observe that the most adopted practices are new practices,  specifically designed for \ac{ML}, followed by traditional and modified practices.
Traditional practices in the ``Team'' category are ranked highly, since collaborative development platforms have become common tools among practitioners and offer good support for information sharing and coordination inside a team.
In contrast, traditional practices related to code quality, such as running regression tests (practice 17) or using static analysis tools to check code quality (practice 19), have low adoption.

\subsubsection*{\textbf{Influence of data type on practice adoption.}}
\label{subsec:influence}

The practices presented in Table~\ref{tbl:practices} are general and should apply to any context.
However, the type of data being processed influences the choice of algorithms and might also influence the adoption of practices.
For example, when processing images or text, it is common to rely on \acp{DNN}, where training is not preceded by a feature extraction step.
Conversely, for other types of ML algorithms, a feature extraction step is common.
Here, we investigate the influence of the type of data to be processed on the adoption of practices.
Moreover, we explore the practices that have distinct adoption rates for specific data types.

The percentage of respondents per data type and the corresponding overall practice adoption rates are presented in Table~\ref{tbl:adoption_data}.
We employ the same percentages described earlier to assess the practice adoption rates per data type.
We observe that, in our data set, tabular data, text, images and videos are predominant (each above $25$\%) and have very similar adoption rates.
Audio and time series have lower representation (under $8$\%), making their adoption rates less reliable. Still, apart from the ``At least high'' category, adoption rates remain similar.  The ``Graphs'' data type is used rarely ($0.5$\%), making adoption rates too unreliable to report.

When comparing the adoption of individual practices, grouped by  data type, we observed that several practices tend to have higher adoption for particular data types.
For all comparisons, we used the ``at least high'' adoption rate.
Firstly, practice 13, on automatic hyper-parameter optimisation, has an adoption rate that is more than 8\% higher for tabular data than for text or images.
This result could be due to the the algorithms or tools used.
The tool support for automatic hyper-parameter optimisation in more traditional ML methods, such as random forests or SVMs~--~which are popular for tabular data~--~is more mature than for newer techniques, \eg,~\acp{DNN}.
Secondly, practice 29, on enforcing privacy and fairness, has an adoption rate for tabular data that is more than 10\% higher than that for text or images.
Lastly, practice 12, on the capacity to run training experiments in parallel, has adoption rates for text and images that are over 10\% higher than that for tabular data.
Perhaps the infrastructure needed to run experiments with text or images~--~where \acp{DNN} are  used extensively and parallelisation is required to achieve good results~--~makes it easier to adopt this practice.

\section{Analysis of Practices and Effects}
\label{sec:analysis}
Following the practice adoption questions, in the questionnaire there were four questions about the perceived effects of adopting these practices.
These questions were designed to test the hypothesis that adopting a set of practices will lead to a desired effect.
A mapping between practices and effects, as hypothesised during survey design, can be found in Table~\ref{tbl:practices_effects}.

\subsubsection*{\textbf{Correlations among practices.}}
\enlargethispage{1\baselineskip}
Firstly, we report results from an ana\-lysis of the correlation between  practices.
We employ the Spearman rank correlation coefficient, $\rho$, in light of the ordinal nature of the Likert scale used in our questionnaire.
In order to determine the statistical significance of the observed correlations, we perform t-tests with a significance level of $0.01$.

In total we found 244 statistically significant, moderate to strong correlations ($\rho\geq0.35$), of which we report on the most informative ones.
For example, writing reusable scripts for data management (practice 3) correlates positively with  testing for skews between different models (practice 23, $\rho=0.35$).
This suggests that the ability to reuse code for data transformation can facilitate model evaluation.
Furthermore, sharing the training objectives within the team (practice 6) correlates positively with using a shared backlog (practice 27, $\rho=0.38$) and using relevant metrics to measure the training objective (practice 7, $\rho=0.43$).
Testing the feature extraction code (practice 8) correlates positively with practices 9 ($\rho=0.35$) and 10 ($\rho=0.56$), on feature documentation and management.
This indicates that practitioners tend to use advanced feature management methods concomitantly and that the feature management practices complement each other.
As expected, practice 8 correlates positively with practice 17, on running regression tests ($\rho=0.37$).

Performing peer review on training scripts (practice 11) correlates positively with all team practices~--~using collaborative development platforms (practice 26, $\rho=0.40$), working against a backlog (practice 27, $\rho=0.44$) and good team communication (practice 28, $\rho=0.44$).
This result is in line with our expectations, since collaborative platforms provide features for code review, and this is further enhanced by good communication within the team.
Peer review also correlates positively with using static analysis tools for code quality (practice 19, $\rho=0.48$), which suggests that teams prioritising code quality apply various techniques for this purpose.

The practices for deployment correlate positively between themselves, suggesting that teams with advanced deployment infrastructures tend to adopt all practices.
For example, automated model deployment (practice 20) correlates positively with shadow deployment (practice 22, $\rho=0.48$) and automated roll backs (practice 24, $\rho=0.51$).
Moreover, continuous monitoring of deployed models (practice 21) correlates positively with logging predictions in production (practice 25, $\rho=0.51$).
These results indicate that the deployment practices are complementary and that adopting some enables the adoption of others.

\subsubsection*{\textbf{Linear relationship between practices and effects.}}
Secondly, we used the initial mapping from practices to effects (presented in Table~\ref{tbl:practices_effects}) to investigate the hypothesis that adopting a set of practices leads to each desired effect.
For the analysis, we trained four simple, linear regression models, one for each set of practices and effects in Table~\ref{tbl:practices_effects}.
For each model, we used the F-test for linear regression to test the null hypothesis that none of the practices are significant in determining the effect, with a significance level of $0.01$.
Since some of the data sets were imbalanced, \ie,~contained substantially more examples for the positive or negative effect, we applied random under-sampling to balance those sets.

The null hypothesis was rejected for all effects; the respective p-values from the F-test are shown in Table~\ref{tbl:practices_effects}.
We also performed t-tests to assess whether any of the coefficients in the regression models were statistically significantly different from zero, and found evidence that (at significance level 0.01) this was the case.
For example, the t-value of practice 25 for traceability is 6.29.
Moreover, the $R^2$ values, also shown in the table, are high for all effects, which indicates that the observed effects are rather well described by a linear model of the degree of adoption of the associated practices.


\subsubsection*{\textbf{Non-linear relationship between practices and effects.}}
Lastly, we report the results from training statistical models to predict each perceived effect from sets of practices.
Unlike the linear regression models described earlier, here, we additionally considered  \ac{ML} models that do not assume a linear relationship between the practices and effects.
Moreover, in order to strengthen the evaluation, we performed hold-out testing, using a test set of 25\% of the data for each effect, which was only used for the final assessment of our models.
We also revised the sets of practices associated with two of the effects (agility and traceability), in order to enhance the prediction accuracy of the models as assessed on validation data. 

Following practice 13 from Table~\ref{tbl:practices}, we considered four types of models with increasing sophistication: (1) simple linear regression models, (2) \ac{RF} regression models resulting from manual hyper-parameter and feature engineering, (3) \ac{RF} regression models whose hyper-parameters were
optimised using grid search, and (4) models obtained from an AutoML system that performed automatic model selection and hyper-parameter optimisation~\cite{NIPS2015_5872}.

During training, we used 5-fold cross-validation on the training data (\ie, the 75\% of the data retained after setting aside the test sets).
For all experiments we used under-sampling on the training data to remove class imbalance.
We also experimented with the SMOTE over-sampling algorithm for regression~\cite{chawla2002smote,torgo2013smote}, but did not observe significant increases in performance of our models.
For the grid search used for hyper-parameter optimisation of our \ac{RF} models, we used 384 candidate configurations for each of the five folds.
For our AutoML approach, we used auto-sklearn~\cite{NIPS2015_5872} with a relatively modest configuration time budget of ~\automltime~and 5-fold cross-validation for internal evaluation of candidate models.

The performance of our predictive models on test data is shown in Table~\ref{tbl:ml_regression}.
For all effects, we used three standard evaluation metrics: mean squared error (MSE), the $R^2$ coefficient of determination, and the Spearman correlation coefficient ($\rho$) for  predicted \emph{vs} true outcomes.
%
We observe that, in all scenarios, the effects can be predicted from the practices with very low error and a high coefficient of determination.
Moreover, the \ac{RF} models always outperform linear regression. This clearly indicates that at least some practices have non-linear impact on the effects we studied.

Models created using AutoML yielded the highest accuracy for two of the effects.
For the two other effects, we observed slight overfitting to the training data for the models obtained from AutoML and hyper-parameter-optimised \acp{RF}.
Nonetheless, the Spearman rank correlation between  predicted and true outcomes is consistently high ($\rho \geq 0.90$) across all models, except those obtained from linear regression.
This indicates that, in all cases, the effects studied can be accurately predicted from the associated sets of practices.

\begin{figure}[t]
    \includegraphics[width=8cm, keepaspectratio]{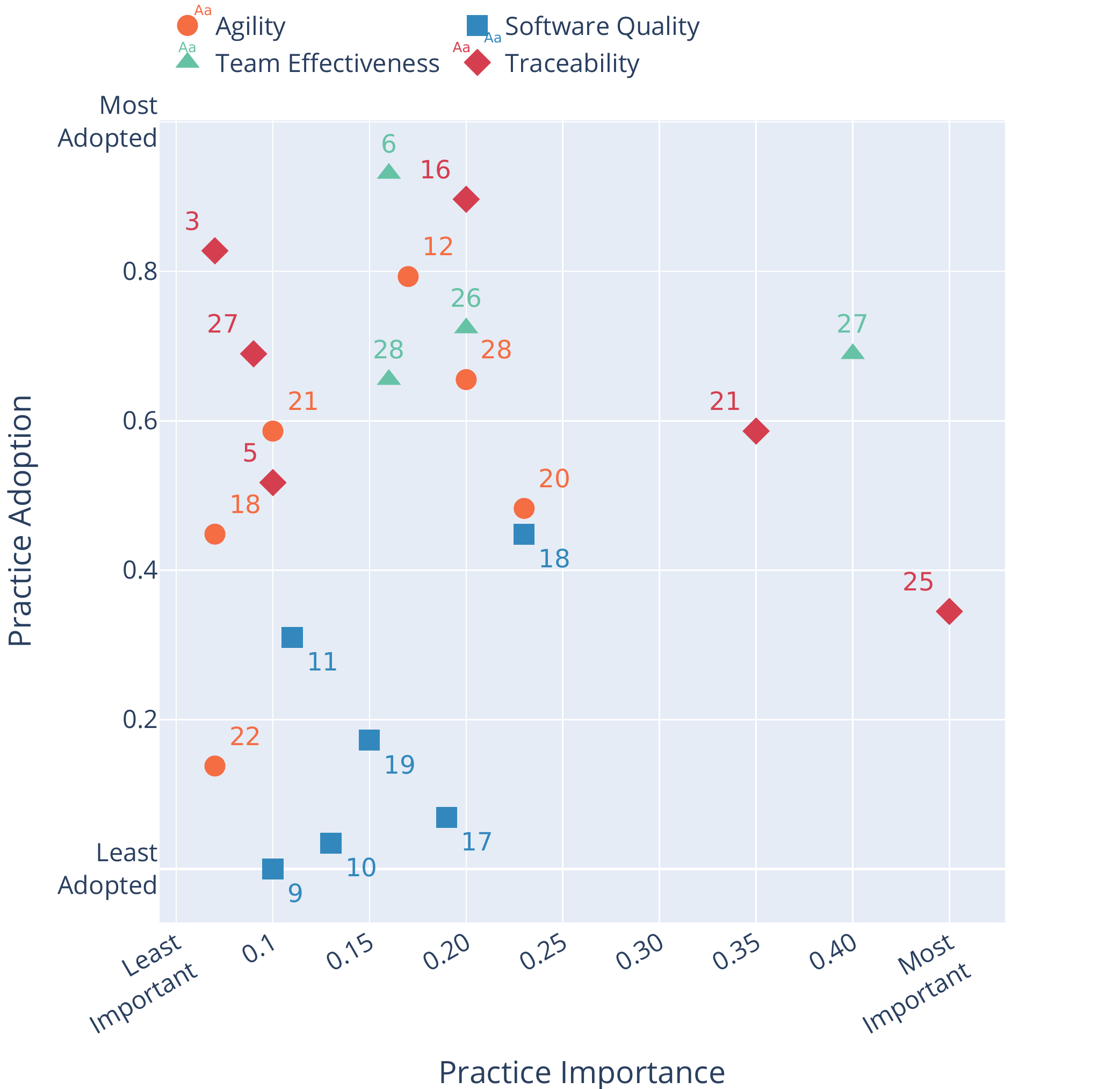}
     \caption{
     Practice adoption and importance, for each effect and practice.
     The practice importance is the Shapley value extracted from the grid search RF models in Table~\ref{tbl:ml_regression}, using the test data set.}
     \label{fig:adoption_importance}
\end{figure}

\subsubsection*{\textbf{Importance of practices.}}
We also studied the contribution of each practice to the final effect, in order to determine the practices that are the most important for each effect.
Towards this end, we used a well-known concept from cooperative game theory called the \emph{Shapley value} to quantify the contributions of individual practices~\cite{FreEtAl16,lundberg2020local}.
In our case, the Shapley value intuitively reflects the increase in predictive accuracy caused by a single practice, averaged over all possible subsets of practices already considered in a given model.
In order to maintain consistency across all effects, and because the models obtained from AutoML are ensembles that are more difficult to analyse, we performed all Shapley value computations for the hyper-parameter-optimised  \ac{RF} algorithms.
We have computed Shapley values on training and test data and obtained consistent results for all effects.

In order to showcase the importance of each practice for an effect, we contrast it with the adoption ranking of the practices from Section~\ref{sec:results}.
We plot the Shapley values and the normalised ranks in~Figure~\ref{fig:adoption_importance}.
The plot indicates, given our data, which practices are most important for achieving a desired effect.
We observe that some very important practices have low adoption rates, while  some less important practices have high adoption rates.
For example, practice 25 is very important for ``Traceability", yet relatively weakly adopted.
We expect that the results from this type of analysis can, in the future, provide useful guidance for practitioners in terms of aiding them to assess their rate of adoption for each practice
and to create roadmaps for improving their processes.
We note that our analysis currently does not take into account functional dependencies between the practices.

\section{Discussion}
\label{sec:discussion}


We now comment on the relation between practice adoption and the challenges from Section~\ref{sec:background}, and discuss threats to the validity of our results. 

\subsubsection*{\textbf{Engineering challenges vs. practice support.}}
When comparing practice adoption (Table~\ref{tbl:practices}) with the engineering challenges referenced in Section~\ref{sec:background}, we observe that many challenges are supported by well adopted engineering practices.

In particular, versioning the artefacts related to \ac{ML} projects, considered a challenge by~\cite{arpteg2018software} and corresponding to practice 16 in our study, has a high adoption rate (rank 3).
The challenges raised by experiment management~\cite{arpteg2018software} and prototyping~\cite{lwakatare2019taxonomy}, such as clearly specifying desired outcomes or formulating a problem (practices 6, 7), as well as monitoring experiments and sharing their outcomes (practices 14, 15), also have high adoption rates.
These results suggest that these challenges have been met by practitioners. 

In contrast, the challenge of testing \ac{ML} artefacts~\cite{arpteg2018software, lwakatare2019taxonomy, ishikawa2019engineers}, corresponds to practices 8 and 17, which have low adoption in our study.
Although we do not detail all testing methods for \ac{ML}, as done in~\cite{zhang2020machine}, the adoption rates for the two testing practices in our study suggests that testing remains challenging.

Several practices presented in this study have low adoption and are not mentioned in previous studies as challenging; this is particularly the case for the practices related to feature management (practices 8, 9 and 10) as well as automating hyper-parameter optimisation and model selection (practice 13).
Although these practices have been recommended in the literature, we plan to further validate their relevance through future participant validation (member check) interviews and by collecting additional data.

\subsubsection*{\textbf{Threats to validity.}}
We identify three potential threats to the validity of our study and its results.
Firstly, the data extracted from literature may be subject to bias.
To limit this bias, several authors with different backgrounds have been involved in the extraction process. 
Also, the pilot interviews and survey produced no evidence suggesting that any of the practices we identified are not recognised by practitioners, nor did we find any indications that important practices were missing from our list. Nevertheless, in the future, we intend to further test completeness and soundness of our catalogue of practices through participant validation interviews.

Secondly, the survey answers may be subject to bias.
As shown in Section~\ref{subsec:demographics}, some groups of respondents are over-represented and may introduce selection bias.
In particular, although the adoption rates for respondents in Europe do not present striking differences when compared to those in South America or Asia, Europe remains over-represented.
Also, some bias may stem from respondents in North America, for which the adoption patterns are different, while they are not equally represented to other groups.
This bias can be removed by gathering more data, as we plan to do in the future.

Lastly, the measurements used to investigate the relationship between groups of practices and their intended effects may be subject to bias.
Rather than measurements of \emph{actual} effects, we used the \emph{perceived} effects as evaluated by the survey respondents.
%
We have not established that perceived effects indeed reflect actual effects, which is an important and  ambitious topic for future research.

\section{Conclusions and Future Research}
\label{sec:conclusion}

We studied how teams develop, deploy and maintain software solutions that involve \ac{ML} components.
For this, we mined both academic and grey literature and compiled a catalogue of 29 \ac{SE} best practices for \ac{ML}, grouped into 6 categories.
Through a survey with \finalFiltered~respondents, we measured the adoption of these practices as well as their perceived effects.
%

\subsubsection*{\textbf{Contributions}}
We reported on the demographic characteristics of respondents and the degree of adoption of (sets of) practices per characteristic.
For example, we found that larger teams tend to adopt more practices, and that traditional \ac{SE} practices tend to have lower adoption than practices specific to \ac{ML}.
%
We also found that tech companies have higher adoption of practices than non-tech companies, governmental organisations or research labs.

Further analysis revealed that specific sets of practices correlate positively with effects such as traceability, software quality, agility and team effectiveness. 
We were able to train predictive models that can predict, with high accuracy, these perceived effects from practice adoption.

We contrasted the importance of practices, \ie, their impact on desirable effects as revealed by these predictive models, with practice adoption, and thus indicating which practices merit more (or less) attention from the \ac{ML} community.
For example, our results suggest that traceability would benefit most from increased adoption of practice 25, the logging of production predictions with model versions and input data.
At the level of teams or organisations, these same results can be used to critically assess current use of practices 
and to prioritise practice adoption based on desired effects.
For example, a team with a strong need for \emph{agility} and low adoption of associated practices may plan to increase  adoption of  those practices.



\subsubsection*{\textbf{Future work}}
We plan to further increase the number of respondents of our survey, so we can perform even more fine-grained analyses.
We may also add more questions, for example to better measure the effects of practices related to AutoML, a 
relatively new direction that is receiving sharply increasing attention in academia and industry. 
We also plan to better cover the traditional best practices from \ac{SE}, using a process similar to the other practices.
Through validation interviews with respondents, we plan to add depth to the interpretation of our findings, especially regarding the relationships between practices and their effects.
We also intend to develop and test a data-driven assessment instrument for \ac{ML} teams, to assess and plan their adoption of engineering practices.
While our study is restricted to \ac{ML} we may also investigate to which extent our findings are applicable for other domains within the broader field of \ac{AI}.
Overall, our hope is that this line of work can facilitate the effective adoption of solid engineering practices in the development, deployment and use of software with \ac{ML} components, and thereby more generally contribute to the quality of \ac{AI} systems.
Furthermore, we are convinced that other areas of AI would benefit from increased attention to and adoption of such practices.


\bibliographystyle{ACM-Reference-Format}
\bibliography{main}

\end{document}